# Full Virtualization of Renault's Engine Management Software and Application to System Development

Yohan Jordan, Dirk von Wissel
Renault SAS - Centre Technique Lardy
1, Allée Cornuel, 91510 Lardy – France

Adrian Dolha, Jakob Mauss
QTronic GmbH
Alt-Moabit 92, 10559 Berlin – Germany

**Abstract**
Virtualization allows the simulation of automotive ECUs on a Windows PC executing in closed-loop with a vehicle simulation model. This approach enables to move certain development tasks from road or test rigs and HiL (Hardware in the loop) to PCs, where they can often be performed faster and cheaper. Renault has recently established such a virtualization process for powertrain control software based on Simulink models. If the number of runnables exceeds a threshold (about 1500) the execution of the virtual ECU is no longer straight forward and specific techniques are required. This paper describes the motivation behind a Simulink model based process, the virtualization process and applications of the resulting virtual ECUs.

*Domain*: Critical Transportation Systems
*Topic*: Processes, methods and tools, in particular: virtual engineering and simulation
*Keywords*: powertrain control, software development, frontloading, validation and test, virtual ECU

## 1. Motivation

Since 2010, Renault has established a framework to develop engine control software for Diesel and Gasoline engines [6]. The framework is heavily based on MATLAB/Simulink and the idea of model-based development, which facilitates the carry-over and carry-across of application software between software projects. In the Renault EMS architecture software is composed in to about 20 *functions*, such as Air System, Combustion etc. A function consists of *modules*. A module is the smallest testable software unit and contains *runnables* to be scheduled and executed by the Operating System (Os) of the ECU. The Renault EMS development process includes basically the following steps [5].

1. Specification of about 200 generic configurable modules per ECU using MATLAB/Simulink.
2. Generation of C code (EMS application software) from all module specifications using MATLAB/Simulink Embedded Coder.
3. MiL (Model in the Loop) test and validation of the resulting executable specifications at module level in a simulated system environment, considering only essential interactions with other modules and system environment. This is essentially a back-to-back test to make sure that the Simulink model of a module and the corresponding production C code show equivalent and intended behaviour. To insure software quality, this step is repeatedly performed with steps 1 and 2, based on the simulation capabilities of MATLAB/Simulink.
4. Configuration of modules to fit to the specific needs of a software project, such as absence or presence of certain components.
5. Integration of generated configured C code and hand-coded platform software (basic software) on supplied target hardware, a real ECU that communicates with other controllers via CAN and other busses.
6. Validation and test of all modules on system level using the real ECU. In contrast to step 3, the interactions of all modules and interactions with the system environment are visible then and subject to testing. For example, the Os runs all scheduled runnables then, not just those of the modules considered to be 'essential' for a module under test.

Critical assessment of the above process shows that there is a considerable delay between delivery of a set of specifications to the software project team (at the end of step 3) and system-level tests based on an ECU that runs entire software (step 6). Typical delays are weeks or months.

Consequently, module developers get feedback about the behaviour of their modules late, i.e., weeks or months after releasing a specification (cf. Figure 1).

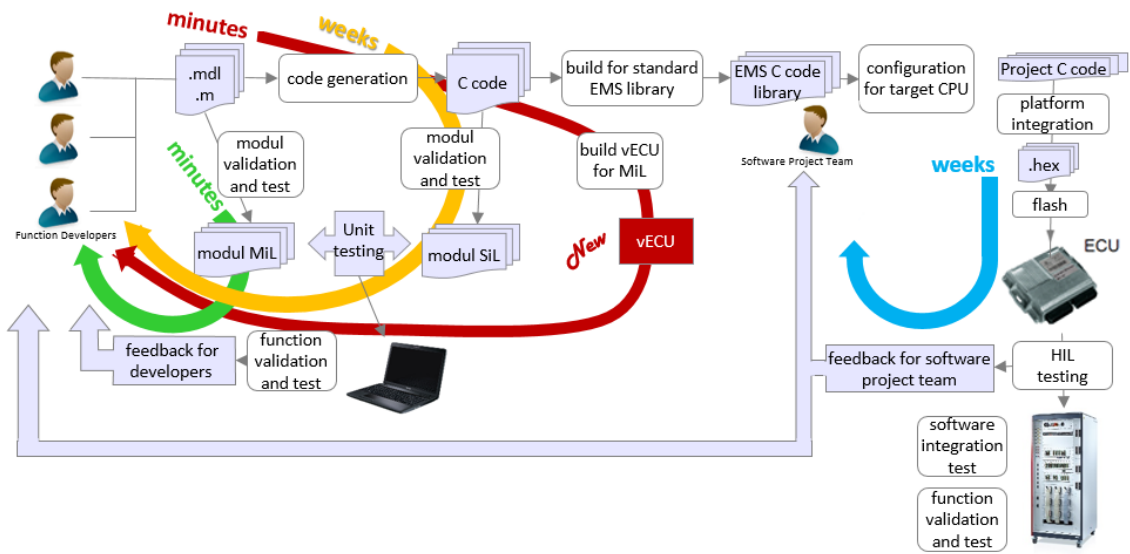

*Fig. 1: Virtual ECU used to frontload system-level validation and test.*

The cycle can be considerable shortened by frontloading system-level validation and test. Ideally, system-level test and validation should be performed interleaved with steps 1, 2 and 3 replacing the actual MiL, SiL validation process by a full ECU validation which would provide the software project team with the possibility to validate integrated software within minutes. Technically, such a frontloading of test activities can be achieved using a virtual ECU, shown as "New" in Figure 1: specifications are used to build a virtual ECU for a model in the loop validation, before production code is generated and integrated with real ECU hardware. This way, the software integration test cycle can then be shortened from weeks and months to minutes.

## 2. Virtualization of the Renault EMS

Motivated by the above analysis, Renault decided in 2014 to complement the established process for EMS development by the ability to generate virtual ECUs early, as shown in Fig. 2. This new feature enables the validation and test of all embedded functions before actual generation of industrial C code. Code generation takes long because it includes several time consuming validations. Right from the start, the intention was to build a virtual ECU based on executable specifications (Simulink models) of a software project to eliminate the delay introduced by the code generation step. To implement this idea, Renault invited leading providers of tools for the virtualization of ECUs to participate in a benchmark. Participants were asked to use their tool to setup an incremental process for building a virtual ECU from given specifications. Main selection criteria were: time needed to configure an initial setup, time needed for incremental rebuild of a virtual ECU, execution speed of the resulting virtual ECU. In this process, the virtual ECU tool Silver (by QTronic, see [1,2,3]) was selected to extend the Renault's existing code developing processes by the ability to work with virtual ECUs."

### 2.1 Incremental build process for a virtual ECU

The Renault EMS is decomposed into about 200 re-usable configurable modules. A module contains so-called runnables. A runnable is specified by a Simulink block that is executed by the operating system (Os) at certain events: periodically, at given crankshaft positions or in response to driver actions such as ignition switched on. The Os specification maps each event to a list of runnables to be executed in response to the event. This coding follows patterns presented in [9]. To handle variants, the configuration of a module for a software project also allows to eliminate some runnables in the C code during the linking phase.

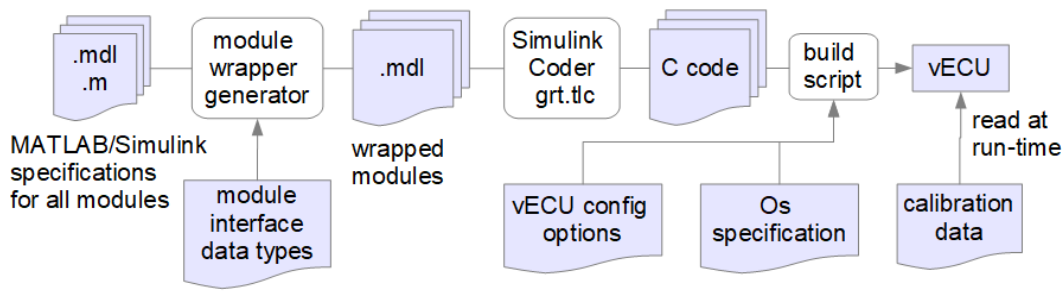

*Fig. 2: Process to build a virtual ECU from given specifications*

The virtualization process is structured as shown in Fig. 2. Starting point is a consistent set of modules, i.e. executable specifications given as Simulink models and MATLAB m scripts, a dictionary that defines data types for all variables exchanged between modules, and an Os specification. Based on this, an incremental, fully automated build process for a virtual ECU is implemented as follows: first, time-stamp information available from the file system is used to identify all modules whose specification has been updated since the previous build. For each such module, a wrapper model is generated that implements the interface of the module to other modules and the Os. The wrapper is a Simulink model and contains all runnables of the corresponding module plus some wiring to enable the Os to call the runnables and to establish signal flow with other modules. In a next step the generated wrapper and the wrapped module is turned into C code using Simulink coder. For this step, the General Realtime Target (`grt.tlc`) shipped with Simulink Coder is used without modification. The C code generated for the module is then post-processed to support on-line calibration. Finally, the C code of all modules of the EMS is compiled together with an Os generated from the given Os specification. The resulting binary runs on Windows PC and is available either as MATLAB/Simulink Sfunction (`.mexw32` or `.mexw64`), as FMU for Co-Simulation [4] or as a binary (Windows DLL) for execution in Silver.

The above build process is controlled by three configuration files to be provided to the build script
- Module interface data types: A dictionary that defines a data type such as `float`, `int16`, `uint8` etc. for each input and output of the wrapped module. To generate C code, Simulink actually determines data types of signals dynamically via propagation through the block diagram. This can lead to conflicting data types when generating C code for each module separately, as we do here. To prevent this, we need to define signal data types statically. One task of the wrapper is to fix the data types of inputs and outputs to the ones given in the dictionary. A user does not need to list all inputs and outputs of a module. This would be cumbersome, because interfaces are frequently changing. Instead, the wrapper generator derives the list of inputs and output automatically by inspecting the wrapped module.
- vECU config options: This selects configuration options to adapt the vECU to specific use cases. A user can specify which additional inputs and outputs the vECU should expose at top-level. For example, he can mark a module as 'by-passable', to support co-simulation of the vECU with Simulink, with the bypassed module running in Simulink as described in section 4.2. He can also decide to turn all tuneable parameters of a certain module into inputs, to support online-tuning of these parameters, see section 4.3.
- Os specification: this is a table that maps each Os event to a list of runnables. The table is used to generate the Os of the virtual ECU. When the named event occurs, the Os runs the corresponding runnables in the given order. We distinguish periodic events such as "10 ms clock tick" from aperiodic events, such as "ignition on". Periodic events are generated by the Os based on simulation time, while aperiodic events are inputs of the Os, to be generated by the engine- or driver model.

Code generation takes about 40 seconds per module, while compilation of the entire C code for a virtual ECU takes about 90 seconds. This means: when only one module is updated, and incremental rebuild of the entire vECU takes less than 3 minutes. This is fast enough to test and validate an updated module in system context immediately after editing the specification of the module. At runtime, a vECU reads a configuration file that contains calibration data. This means, calibration data can be varied and validated without recompiling the virtual ECU. This way, calibration data becomes part of the validation and test loop as well.

The incremental strategy used to build a vECU allows us to generate C code for a single module independent from all other modules. We also tried the opposite strategy: load all modules of the ECU and a generated model of the Os into Simulink and generate C code in one pass. This proved to be infeasible because initialization of the resulting model takes very long (> 10 hours) and the code generator tends to run out of memory during subsequent code generation, even with 64-bit Simulink.

### 2.2 Differences between real and virtual ECU

A virtual ECU is a model of the real ECU. Consequently, not all properties of the real ECU are visible in the virtual ECU. This means that not all tests that are relevant can also be moved to the PC. For some tests, access to real ECU hardware is needed.

The most crucial differences between real and virtual ECU as presented here are

- *Zero-time execution*: the virtual ECU behaves like a device with unlimited computing speed. As a consequence, the Os of a virtual ECU cannot interrupt a running task (no pre-emptive multi-tasking, since a task takes no time to execute). All tasks run exactly as scheduled (for example, cyclically with 1, 2 or 10 ms period), no matter how long its execution may take on the real target hardware.
- *Missing basic software*: Hand-coded basic software of the ECU platform (e.g. drivers for Can, Lin, FlexRay) is not part of the virtual ECU and can therefore not be validated. This is not a fundamental restriction of virtual ECUs, but a consequence of the EMS development workflow shown in Fig. 1.
- *Different production code*: C code representation of calibration data and signals as generated with Simulink Coder is close to but not fully equivalent to the corresponding representations of the production code. For example, on both sides, the same scalar data types are used: `float`, `uint8`, `int16` etc. However, C code generated by Simulink Coder for grt represents all signals and tuneable parameters as members of a C struct, while in production C code, these are represented as global variables. Consequently, certain defects cannot be found using the virtual ECU.

Many of these differences can be avoided by switching to the SiL ([1],[3]) or vPiL (virtual processor, [2],[3]) type of virtual ECU. However, as explained above, this was not an option right now, due to the time constant for code generation and workflow issues, but might become feasible in the future.

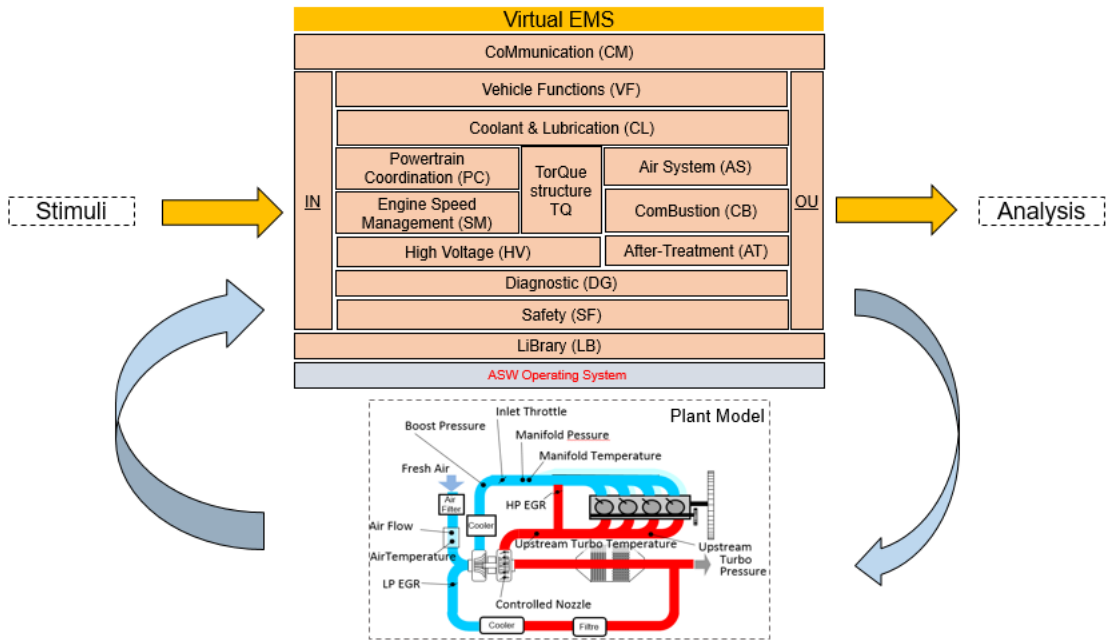

*Fig. 3: Virtual ECU of Renault EMS connected to engine model*

### 2.3 Closed-loop simulation with engine model

In order to study the behaviour of different feedback loops of the EMS, the vECU created by the above process are typically executed in closed-loop with an engine model. Currently, real-time capable engine models developed by Renault using the multi-physics modelling tool LMS Imagine.Lab Amesim (LMS Amesim) for HiL applications are re-used for this purpose. However, an appealing property of virtual ECUs is its complete independence from real time. A virtual ECU can easily be coupled with plant models that run slower or faster than real time. This enables the use of computational expensive plant models, for example in the domain of combustion simulation or exhaust after treatment.

Such models cannot be used in a real-time environment, but without problems in conjunction with virtual ECUs.

### 2.4 Runtime performance of the virtual ECU
A virtual ECU for a typical Renault EMS loads and initializes in less than 5 seconds on a Windows PC. In our experiments, ECU behaviour was sampled with 1 ms steps. Execution speed of the vECU on PC depends on the number of variables (outputs of the vECU) to be recorded during simulation.
- When measuring 170 variables per ms, execution of the vECU is 4 times faster than real time.
- When measuring 20.000 variables per ms, execution speed is 3 times slower than real time.

The engine model used has been developed for HiL applications and is therefore very fast. In a closed-loop simulation with a virtual ECU on PC, the engine model consumes less than 10% of the computing time.

## 3. Virtual ECUs: Related Work
In this section, we review existing work on the virtualization of ECUs and motivate the approach chosen here. As shown in Fig. 4, we distinguish three options to create a virtual ECU for execution on a PC.

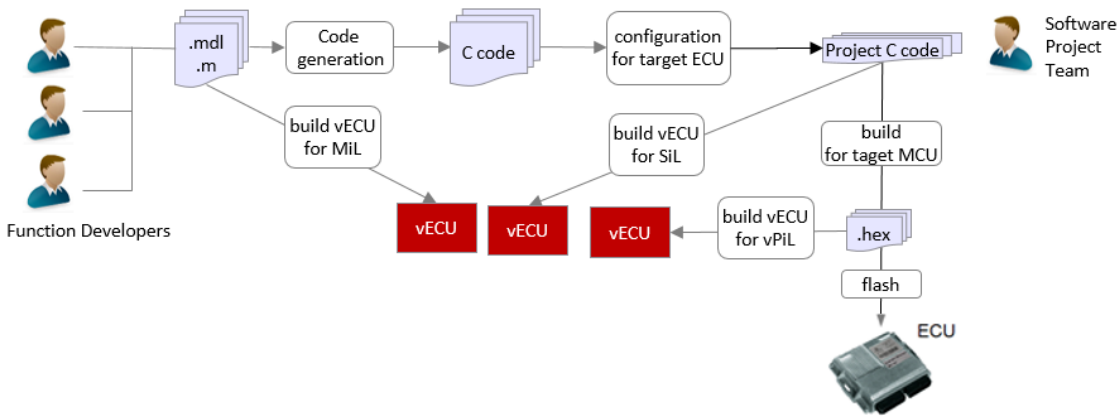

*Fig. 4: Virtual ECU for MiL, SiL and virtual PiL*

Depending on the use case to be supported, a virtual ECU is created from either
- virtual PiL: the hex file resulting from compiling C code for the target processor
- SiL: C code compiled for PC
- MiL: models that generate the C code

### 3.1 Virtual PiL: Chip simulation to run a hex file on PC
At the heart of an engine control unit (ECU) is a microcontroller unit (MCU), i.e. a digital processor with additional peripherals such as CAN controllers, analog digital converters (ADCs) etc. integrated on a single chip. Wide spread MCU families for engine control are Tricore (Infineon), PowerPC (Freescale/NXP, STM), and v850 (Renesas). MCUs for engine control often have two or more cores for fast concurrent computing. A virtual ECU can be created by simulating the corresponding MCU. The hex file resulting from compiling the EMS software for the target MCU contains binary program code and calibration data. For example, for a Tricore MCU, the four bytes `0x50, 0x1F, 0xF1, 0x8B` (given in hexadecimal notation) encode the assembler instruction

```
add %d5, %d1 -1
```

which instructs the MCU to take the 32-bit integer currently stored in data register d1, decrement the value by one and store the result in data register d5. A chip simulator reads instructions like this from simulated memory, decodes these and performs the corresponding computation, which may update register and memory content. Chip simulators come in different flavours: A chip simulator is *instruction accurate*, if it correctly simulates the entire instruction set of the target MCU. A chip simulator is *cycle accurate*, if the simulator also correctly predicts how many clock ticks it takes the MCU to run an instruction. Instruction accurate simulators at typically an order of magnitude faster than cycle accurate simulators because a cycle accurate simulator requires a far more detailed model of the underlying MCU, to take into account effects of multi-core processing, instruction pipelines, caching strategy, and speed of memory access on processor speed. Besides, chip simulators differ in the coverage of their chip model: An *instruction set simulator* just simulates the instruction set of the MCU, while a *platform simulator* covers also on-chip peripherals integrated on the chip, and maybe even ASICs and other MCU external chips found in the ECU. A pure instruction set simulator can be evolved into a platform simulator by integrating additional models, typically using some kind of plugin API provided by the simulator.

Cycle accurate platform simulators can be implemented using hardware description languages such as SystemC or VHDL, often based on design data (IP) owned by the chip maker, while instruction set simulators are typically implemented based on a description of the corresponding instruction set.
In the context of engine control (and beyond), there are two major use cases for chip simulation
- Suppliers of ECU hardware and MCUs use chip simulators to validate their ECU and chip designs, for example w.r.t computing performance. Such analysis typically requires a cycle accurate platform simulator as e.g. provided by Synopsys.
- OEMs that need to integrate and calibrate supplied EMS software use chip simulators to run the EMS software on PC in cases where they have access to the hex file, but no access to the C code of the EMS. Fast instruction set simulators are often used for this type of application. For example, [2, 3] reports about simulation of an engine controller using Silver's chip simulator. On a typical PC, a modern ECU runs then approximately in real-time.

**3.2 SiL: Compiling the C Code for PC**
In many application scenarios, the C code of the ECU to virtualize is available. A virtual ECU can then be created by compiling the C code for PC. Compared to chip simulation, the resulting virtual ECU runs typically faster and provides better support for C level debugging, if compiled with debug option. This is important, because testing and debugging of EMS code on PC is a mayor use case for virtual ECUs. Typically, only the modules of the application software and optionally selected modules of the basic software are compiled for PC, but not the operating system (Os) and hardware drivers, such as the AUTOSAR MCAL layer. Instead, the services provided by Os and drivers are re-implemented for the PC environment. In the virtual ECU, these PC-specific implementations replace their corresponding MCU-specific counterparts.

Some tools used to create virtual ECUs based on given C code are EVE (ETAS), Silver (QTronic, [1, 2, 3]) and VEOS (dSPACE). Available tools differ w.r.t. the build process for vECUs, their handling of AUTOSAR and non-AUTOSAR projects, their ability to process standardized ECU and bus descriptions (such as ASAP2 `.a2l`, AUTOSAR `.arxml`, CAN `.dbc`, LIN `.ldf`, and FlexRay/Fibex `.xml`), treatment of Os and drivers and runtime support for debugging, scripting, measurement and calibration. Good support for the above descriptions enables to automate the creation of virtual ECUs.

In principle, a ECU can also be virtualized using a hypervisor as discussed in [10]. In our case, this would require to run the Os of the EMS ('guest Os') on Windows ('host Os').

**3.3 MiL: Run the models used to generate C Code**
Today, model-based development (MBD) on PC with tools like MATLAB/Simulink (MathWorks), ASCET (ETAS), SCADE (Ansys) or Rational Rhapsody (IBM) is the established method of choice for the development of control software. For automotive applications, Simulink seems to dominate the market.

Surprisingly, the dominance of MBD does not mean that developers are typically able to simulate their ECU on PC, even if they have full access to all models of the ECU. For example, loading and initializing all 200 modules of a typical Renault EMS into 64-bit Simulink takes about 10 hours. As far as we can see, the delay is mostly caused by the run-time propagation of data types in the model initialization phase. Interactive simulation of a module in full ECU context is clearly out of reach then. The only way to achieve reasonable execution times is to apply compilation techniques, either within the Simulink environment, or (as we did here), in a Simulink external integration environment, based on exported C code. As the case study presented here shows, such a compiled environment is not straight forward to set up. In our case, it took many months. We could also not find reports about similar attempts in the literature. It seems that developers have widely accepted the fact, that they cannot run the ECU model inside their model-based IDE. Instead, they have to wait until the C code generated by the models has been integrated with the target ECU hardware, and the ECU software can be executed on a HiL system. Of course, this practice fundamentally contradicts the idea of MBD which is about executable models whose behaviour can be explored and assessed during design.

### 3.4 Motivation for choosing the MiL type of virtual ECU
The objective of the project reported here has been to improve the quality of executable module specifications by providing a virtual ECU that runs all ECU modules simultaneously, only minutes after a module has been edited. It was therefore a natural choice to build the vECU from the module source (Simulink .mdl files), which lead us to a MiL type of vECU. The production C code generator is a shared and limited resource here, not easily available in the working environment of module developers. This currently rules out the SiL type of vECU. Hex files for the target MCU become only available weeks after a module edit (see Fig. 1), which rules out the vPiL type of vECU.

## 4. Applications of virtual ECUs at Renault engine development
In 2016, Renault created the first fully functional virtual EMS using the process described above. The process has been repeated for about 6 releases (updates and different platforms) of the EMS software since then. Besides, we focus work on setting up the tool chain for the most promising applications of virtual ECUs. These applications are briefly described here, ordered by their location on the time-line of the development process. For each such application, we also sketch the current state of the implementation, if applicable.

### 4.1 Definition of system requirements
In the requirement definition phase of a new function, a virtual ECU can be used to complete an environmental model when interactions between systems through control actions need to be taken into account to reflect the real system behaviour. This way, simulation can be used to derive (even quantify) requirements. We have not yet implemented this idea based on virtual ECUs as described above.

### 4.2 Module development in system context
As described above, a main use case for virtual ECUs is validation and test of module specifications in system context before the specifications are forwarded to the software project team. Such tests are best performed from within the development environment (IDE) for modules, which is Simulink here. To support this, virtual ECUs can be configured at compile time for co-simulation with Simulink. In the vECU configuration dialog, a module developer can select his or her module for execution in Simulink, bypassing (replacing) the implementation of the module provided by the virtual ECU. The module developer can then edit his module in Simulink, for example modify the Simulink block that implements a certain runnable. After such an edit, the developer can immediately run the virtual ECU, without rebuilding it. This runs all runnables of all modules of the vECU, except those of the bypassed module, which run in Simulink then. This way, the effect of an edit on behaviour can immediately be seen by the developer in the context of all other modules of the ECU. In such a co-simulation scenario, rebuild of the vECU is only required when the Os specification or the signal interface for the edited module changes, or to get more recent implementations of other modules into the validation loop. Thanks to incremental build, such a rebuild takes only minutes.

### 4.3 Pre calibration of the EMS
At least 50% of the development time for engine controllers is spent for engine calibration, i.e. tuning of software parameters of the EMS in order to simultaneously meet customer demands (for example, 'fun to drive'), as well as legal requirements concerning emissions and fuel consumption. With virtual ECUs, Renault started to frontload calibration related activities as well. This is called pre-calibration. The objective is to develop a better starting point for calibration early, in order to gain more time for fine tuning in later phases of development. As pointed out before, a Silver virtual ECU can be configured to load calibration data at runtime from a human readable text file. This means that calibration data can be varied without rebuilding the vECU, which speeds up the calibration process. To support interactive on-line tuning of parameters, a virtual ECU can also be configured to expose all tuneable parameters of selected modules as permanent (dynamic) inputs of the vECU. This way, a calibration engineer can vary parameters even during simulation, typically using a slider provided by Silver's user interface.

### 4.4 Virtual integration of modules before production C code is generated
Virtual ECUs enable us to virtually integrate all modules of an EMS and to test the resulting virtual ECU in closed loop with an engine simulation before turning the modules into production C code. This helps to shorten development cycles. Integration problems can now be detected more early, before integrating production C code on real ECU hardware. To exploit this idea, we are currently comparing simulation results recorded on a HiL test system with simulation results computed on PC using a virtual ECU for sufficiently similar versions of EMS software and calibration data. Unexpected differences quickly point us to potential problems.

### 4.5 Move selected tests from HiL to MiL
Many of the engine tests that run today on limited resources like HiL test systems, engine test rigs and real vehicles could actually be moved to highly available PCs, because they do not depend on properties of the underlying hardware platform. In such a process, only certain releases of the EMS software (in particular the final release) are tested using real target hardware, while all other intermediate releases are tested on virtual platforms. This can for example be used to increase test coverage. Virtual ECUs are a key tool to implement such a mixed strategy. This idea has not yet been explored using the virtual ECUs described here.

### 4.6 Support joint development within the Renault-Nissan partnership
Renault and Nissan jointly develop certain EMS modules. The development of a shared set of reusable EMS modules requires rework of interfaces and methodologies on both sides. This is a complex process. In this context, Silver virtual ECUs are currently used as a platform for virtual integration of modules. This helps to shorten development cycles and to discover and eliminate problems more early.

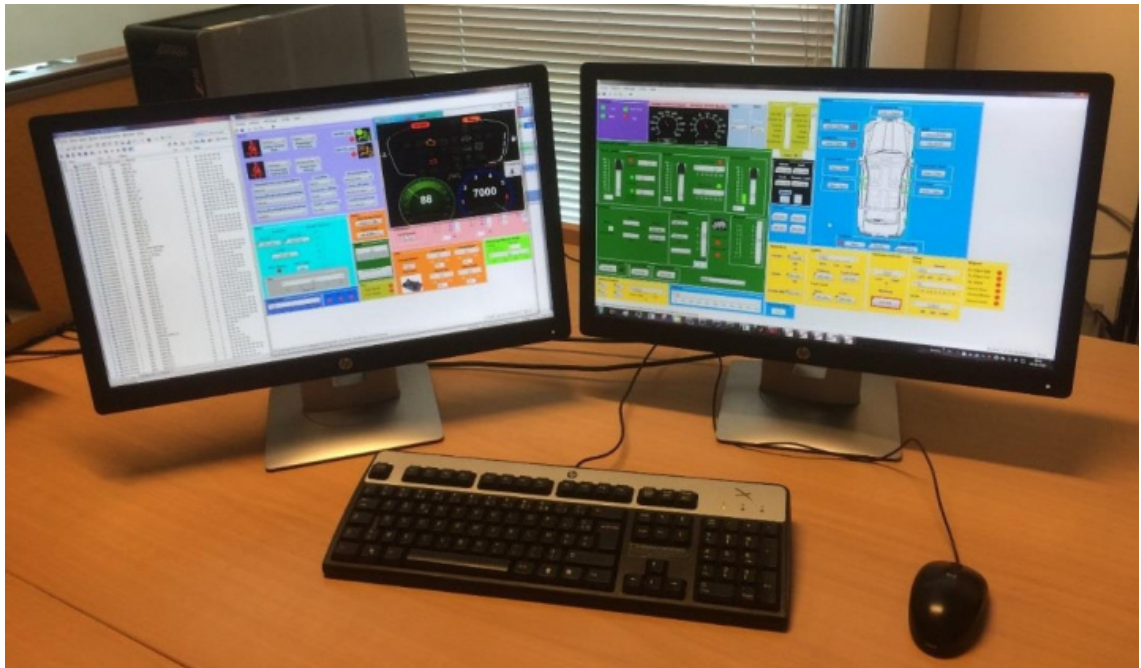

*Fig. 5: D-EIPF running a full vehicle simulation*

### 4.7 Improve existing vehicle-level simulation
Renault and Nissan use the Digital Electronic Integration PlatForm (D-EIPF, see [8]) to validate networked ECUs in vehicle context. With D-EIPF, the communication behaviour of networked ECUs of an entire vehicle can be simulated without the need to access real ECU or other vehicle hardware.
Since 2010, D-EIPF has been developed in-house on top of the RT-LAB Orchestra API (OPAL RT). Network simulation is based on CANoe (Vector). Silver has been integrated into the D-EIPF environment. This way much more realistic engine models (i.e. a virtual ECU running in closed loop with an engine plant model) can now be used to validate overall vehicle behaviour, which increases the scope of tests that can be executed on D-EIPF.

## 5. Conclusion
Renault started to use virtual ECUs to frontload test and calibration related activities during the development of engine management software. First results of these activities are reported in this paper and have been encouraging so far. In the long run, the existing MiL-based virtual ECUs will be complemented by SiL-based virtual ECUs. The latter are based on production C code and will contain basic software as well, which further minimizes the behavioural gap between real and virtual ECUs. This will enable us to move even more development steps to the virtual platform.